\documentclass[%
 superscriptaddress,
preprint,
 amsmath,amssymb,
 aps,showpacs 
]{revtex4-1}

\usepackage{graphicx}
\usepackage{dcolumn}
\usepackage{bm}
\usepackage{color}

\begin{document}


\title{Hamiltonian adaptive resolution simulation for molecular liquids}


\author{Raffaello Potestio}
\affiliation{Max-Planck-Institut f\"ur Polymerforschung, Ackermannweg 10, 55128 Mainz, Germany}
\author{Sebastian Fritsch}
\affiliation{Max-Planck-Institut f\"ur Polymerforschung, Ackermannweg 10, 55128 Mainz, Germany}
\author{Pep Espa\~{n}ol}
\affiliation{Dep.to de F\'{\i}sica Fundamental, Facultad de Ciencias (U.N.E.D.),  Avda. Senda del Rey 9, 28040 Madrid, Spain}
\author{Rafael Delgado-Buscalioni}
\affiliation{Dep.to de F\'{\i}sica Teorica de la Materia Condensada, Universidad Aut\'{o}noma de Madrid, Campus de Cantoblanco, 28049 Madrid, Spain}
\author{Kurt Kremer}
\affiliation{Max-Planck-Institut f\"ur Polymerforschung, Ackermannweg 10, 55128 Mainz, Germany}
\author{Ralf Everaers}
\affiliation{Laboratoire de Physique et Centre Blaise Pascal, {\'E}cole Normale Sup\'erieure de Lyon, CNRS UMR5672, 46 all\'{e}e d'Italie, 69364 Lyon, France}
\author{Davide Donadio}
\email{donadio@mpip-mainz.mpg.de}
\affiliation{Max-Planck-Institut f\"ur Polymerforschung, Ackermannweg 10, 55128 Mainz, Germany}

\begin{abstract}
Adaptive resolution schemes allow the simulation of a molecular fluid treating simultaneously different subregions of the system at different levels of resolution. In this work we present a new scheme formulated in terms of a global Hamiltonian. Within this approach equilibrium states corresponding to well defined statistical ensembles can be generated making use of all standard 
Molecular Dynamics or Monte Carlo methods. Models at different resolutions can thus be coupled, and thermodynamic equilibrium can be modulated keeping each region at desired pressure or density without disrupting the Hamiltonian framework.
\end{abstract}

\pacs{05.10.a, 02.70.Ns, 61.20.Ja, 82.20.Wt}

\maketitle

Complex molecular fluids and soft matter typically display inherently multiscale phenomena and properties. To handle this problem general strategies have been developed, which can be classified as either \textit{sequential} or \textit{simultaneous}. In the former class of methods, coarse-grained models are (usually) developed from microscopic input \cite{kremer2000,vdvegt2009,Hijon2010}; systems are then  simulated separately at different levels of resolution.
In the latter class, which we pursue here, systems are treated within a single simulation on different levels of resolution. A small, well defined region of space is kept at a higher level of detail, while the surrounding can be treated on a coarser, computationally more efficient level.

This idea has been successfully employed, for example, to investigate crack propagation in hard matter \cite{rudd,Rottler:2002,Csanyi:2004,Jiang:2004,kax} and in mixed quantum mechanics/molecular mechanics (QM/MM) simulations, where particles are assigned statically to either the QM or the MM region~\cite{WARSHEL:1976p4538,qmmm1,Svensson:1996p4537,Carloni:2002p4461,bulo}. For soft matter and liquids, inherent fluctuations and particle diffusion require a setup where molecules, or parts of them, can cross boundaries between areas at different resolution, while maintaining the overall thermodynamic equilibrium. Scale-bridging methods have been developed in various fashions to couple all atom (AA) and coarse-grained (CG) models~\cite{annurev}, and even particle-based models to the continuum 
\cite{delgado:2008,delgado:2009,delgado:2006}. To our knowledge, to date the only energy-conserving mixed-resolution approach is the `Adaptive Partitioning of the Lagrangian' method by Heyden and Truhlar \cite{heyden:2008,heyden:2009}. In this method, a combinatoric sum of all possible AA and CG interactions between molecules in different resolution regions is taken into account. The practical viability of this approach is limited by its intrinsic combinatoric complexity, and by the fact that the resulting equations of motion are not amenable to a standard symplectic integrator (e.g. velocity Verlet of leap frog), so that an \textit{ad hoc}, more complicated one had to be developed.

\begin{figure}
\centering
\includegraphics[width=7.5cm]{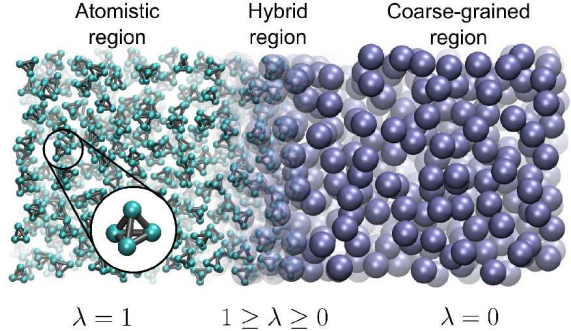}
\caption{Illustration of the simulation setup for adaptive resolution simulations. Molecules freely move from an atomistic region, AA, $\lambda=1$, through a transition zone H to the coarse-grained region, CG, $\lambda=0$. $\lambda(\textbf{R})$, \textbf{R} being the position of the center of mass of the molecules, is a smooth transition function that interpolates between the AA and the CG region.}
\label{setup}
\end{figure}

With  this  idea  of  mixed  resolution in  mind  the AdResS (Adaptive
Resolution Scheme) method was developed, in which one
can  dynamically  couple  specific  regions  of a  simulation  box  at
different  levels   of  resolution,  while   maintaining  the  correct
thermodynamic equilibrium between them
\cite{jcp,adress2,adress3,annurev,jacs1,adresstoluene,adolfoprl,potestio}. The
particles  move  from  one  region  to  the  other  through  a  hybrid
resolution  zone (Fig.  \ref{setup}):  in this  region the  resolution
switch  is defined  by  a transition  function $\lambda(x)$,  smoothly
changing the interactions  from an atomistic description, $\lambda=1$,
to a coarser one, $\lambda=0$, which typically contains a considerably
smaller number of  degrees of freedom (DOF's) per  molecule. AdResS is
based on  the requirement  that molecules interact  through pairwise
forces,  and Newton's  third law  is strictly  satisfied in  the whole
simulation  box by construction.  These requirements  lead to  a force
interpolation      scheme      between      molecules,     $      {\bf
  F}_{\alpha\beta}=\lambda( {\bf R}_\alpha)\lambda({\bf R}_\beta){\bf
  F}_{\alpha\beta}^{AA}+[1-\lambda({\bf R}_\alpha)\lambda({\bf R}_\beta)]{\bf
  F}_{\alpha\beta}^{CG},  $ where  the  force ${\bf  F}_{\alpha\beta}$
between centers of  mass of molecule $\alpha$ and  $\beta$ consists of
an          atomistic,          $\lambda({\bf R}_\alpha)\lambda({\bf R}_\beta){\bf
  F}_{\alpha\beta}^{at}$      and      a     coarse-grained      part,
$[1-\lambda({\bf R}_\alpha)\lambda({\bf R}_\beta)]{\bf
  F}_{\alpha\beta}^{cg}$.
  Yet, it was formally demonstrated \cite{prelu} that a Hamiltonian compatible
  with this force interpolation scheme can not exist.

The method is nonetheless robust, since it allows us to define temperature, pressure
and density everywhere, and  the introduction of a thermodynamic force
\cite{adresstf}  in  the  transition  zone  paved  the  way  to  open  system  MD
simulations  \cite{FritschPRL,jcp}.
Despite the success of the force-interpolation based  AdResS method, though, the lack of a
Hamiltonian  description in the  transition region  is a  drawback: it
hampers a general  statistical theory for the whole  setup, limits the
choice   of  the   simulation  ensemble   and  prevents   Monte  Carlo
simulations.
Moreover, in  the transition  region the
system    has   to    be    stabilized   by    a   local    thermostat that
removes excess heat thus keeping the system in a state of dynamical equilibrium
\cite{jcp,adress2,adress3,annurev,adresstoluene,adolfoprl,potestio,ensing,prlcomment2011}.
In this Letter we propose a new resolution-interpolation and coupling concept, $H$-AdResS, which is formulated in terms of a general Hamiltonian $H$ for the whole system. 
Furthermore,  we  develop  an   analogy  to  the  Kirkwood \cite{kirkwood1935}  coupling
parameter scheme  where we relate  the variation of  the thermodynamic
properties  through the transition  zone in  H-AdResS to  the
integration over $\lambda$ in homogeneous systems. 
We demonstrate our approach on a prototypical mixed AA/CG system, showing that the existence of a global Hamiltonian makes it possible to perform microcanonical (NVE) adaptive resolution simulations.

Let us consider a system composed of $N$ molecules.~\footnote{For simplicity we only consider whole molecules; however, in the same fashion one could consider suitable fragments of a macromolecule.} (labeled by greek indices), each having $n$ atoms (labeled by latin indices) The resulting $M = n N$ atoms interact {\it via} general intramolecular potentials, and pairwise intermolecular potentials. The Hamiltonian of this system can be written as:
\begin{eqnarray}\label{atomistic_H}
&H^{\text{AA}}& = \sum_{\alpha=1}^N \sum_{i=1}^n \frac{p_{\alpha i}^2}{2 m_{\alpha i}} + \sum_{\alpha=1}^{N} V_\alpha^{AA} + V^{int}\\ \nonumber
&V^{\text{AA}}_\alpha& \equiv \frac{1}{2}\sum_{\beta\neq \alpha}^{N} \sum_{ij}^n V^{AA}(|\textbf{r}_{\alpha i} - \textbf{r}_{\beta j}|) \nonumber
\end{eqnarray}
$V^{int}$ indicates the intramolecular interaction, for which we do not need to make any assumption. $p_{\alpha i}$, $m_{\alpha i}$, and $\textbf{r}_{\alpha i}$ are the momentum, mass, and position, respectively, of atom $i$ of molecule $\alpha$.
We now consider a CG pair potential $V^{CG}_{\alpha\beta} \equiv V^{CG}(\textbf{R}_\alpha-\textbf{R}_\beta)$ that depends on the center of mass (CoM) positions $\textbf{R}$ of the molecules $\alpha$ and $\beta$; the total CG potential energy of molecule $\alpha$ is given by $V^{CG}_\alpha \equiv \sum_{\beta\neq \alpha} V^{CG}_{\alpha\beta}/2$. In analogy to Kirkwood's thermodynamic integration (TI) method to compute free energy differences \cite{kirkwood1935}, we define a `mixed resolution' Hamiltonian $H$:
\begin{eqnarray}\label{Hmix}
&&H =\\ \nonumber
&&\sum_{\alpha i} \frac{p_{\alpha i}^2}{2 m_{\alpha i}} + \sum_{\alpha} \left\{\lambda_\alpha V^{AA}_\alpha + (1 - \lambda_\alpha) V^{CG}_\alpha \right\} + V^{int}
\end{eqnarray}
 where coupling parameters $\lambda_\alpha = \lambda(\textbf{R}_\alpha)$
 were introduced, which depend on the position of the molecules' centers of mass.
 The local resolution $\lambda(\textbf{{\bf  R}})$
varies  between $1$  (completely  AA system)  and  $0$ (completely  CG
system);  intermediate  $\lambda$   values  define  a  hybrid  region,
interfacing  the   atomistic  and  the  coarse-grained   ones  (as  in
Fig. \ref{setup}). According to the Hamiltonian in Eq. \ref{Hmix}, for
molecules interacting with a mixed-resolution (having $\lambda \in (0,
1)$)    both    AA     and    CG   \textit{total} potential  energies  are calculated and  weighted according to  their own
resolution  $\lambda$.  The atomistic  DOF's  are retained  everywhere
\cite{jcp}  and   their  dynamics  is   seamlessly  evolved,  allowing
coarse-grained molecules  close to the hybrid region  to interact also
at the atomistic level. Being defined in terms of a Hamiltonian, this hybrid-resolution scheme
conserves the  total energy in  a microcanonical simulation, as  it is
numerically  verified \cite{SI}.   Furthermore,  as different  regions
exchange  particles and energy  the resulting  stationary state  is an
{\em equilibrium state}.

The force derived from $H$ (Eq.  \ref{Hmix}) has the form:
\begin{eqnarray}\label{forces}
&&\textbf{F}_{\alpha i} =\\ \nonumber
&&\sum_{\beta,\beta\neq \alpha} \left\{ \frac{\lambda_\alpha + \lambda_\beta}{2} \sum_{j=1}^n\textbf{F}^{AA}_{\alpha i|\beta j} + \left(1 - \frac{\lambda_\alpha + \lambda_\beta}{2}\right) \textbf{F}^{CG}_{\alpha i|\beta} \right\}\\ \nonumber
&&+\ \textbf{F}^{int}_{\alpha i} - \left[ V^{AA}_\alpha - V^{CG}_\alpha \right] \nabla_{\alpha i}\lambda_\alpha
\end{eqnarray}
where           $\textbf{F}^{AA}_{\alpha          i|\beta          j}$
(resp. $\textbf{F}^{CG}_{\alpha i|\beta}$) is  the AA (resp. CG) force
acting on  atom $i$ of molecule  $\alpha$ due to  the interaction with
molecule $\beta$.  The distribution  of CoM forces  onto the  atoms is
described   in  Ref.~\cite{FritschPRL}.   The first term of Eq. \ref{forces} contains a weighted  sum of
pairwise  forces and is  antisymmetric by  exchange of  the molecules'
labels; this term therefore  complies with Newton's 3rd law everywhere, and is analogous to the AdResS force interpolation.
The second term, $\textbf{F}^{int}_{\alpha i}$, is the
force exerted  on atom  $\alpha i$ by  the other atoms  \textit{in the
  same molecule} and does not  contribute to the force balance between
molecules.   The  third term,  ${\bf  F}^{dr}_{\alpha i} \equiv  -\left[
  V^{AA}_\alpha     -     V^{CG}_\alpha     \right]     \nabla_{\alpha
  i}\lambda(\textbf{R}_\alpha)$,  introduces in  the  hybrid region  a
drift   force   which  violates   Newton's   3rd   law  and   momentum
conservation.
${\bf  F}^{dr}_{\alpha i}$ plays  the role of  an external
force  inducing, in general, pressure and density inhomogeneities in  the  system 
as  it reaches equilibrium (no temperature gradients
are present because energy is conserved and freely flows between
the two subdomains). In particular, the drift force is balanced by a hydrostatic
pressure gradient across the hybrid region given by
$\nabla p = \rho \langle {\bf F}^{dr} \rangle$ \cite{fluid_mech}.

We validated our approach on the same model system as in \cite{jcp} and illustrated in Fig.~\ref{setup}. A detailed description of these simulations is given in the supplementary information~\cite{SI}.
The AA system consists of tetrahedral molecules, each composed by four atoms of unit mass connected by anharmonic springs.
The atomistic interaction between molecules is given by a purely repulsive Weeks Chandler Andersen (WCA) potential, while the CG potential was obtained via Iterative Boltzmann Inversion (IBI)~\cite{reith2003deriving}.
In contrast to the original AdResS scheme, we could perform these adaptive resolution simulations in the microcanonical ensemble, achieving conservation of the total energy (Fig. S1). The resulting density profiles are flat in both the AA and the CG regions, and within $1\%$ of the reference values (Fig. S2). In addition, the structure of the fluid in the AA region is unchanged compared to a purely AA simulation (Fig. S2). 

Such seamless coupling stems from the good matching between the thermodynamic properties of the AA potential and the CG potential. In general, however, the employed CG potentials are only approximations to the  the exact many-body CG potential \cite{gordon2007,wang2009comparative}. As a consequence, there is usually a thermodynamic mismatch between the AA and the CG systems, which can have different chemical potentials and equations of state~\cite{Hijon2010}. If compensations are introduced as time-independent functions of the position of the molecules, then the modified Hamiltonian $\widehat H$ is expressed as
%
%
\begin{equation}
\label{Hmix2}
\widehat H = H - \sum_{\alpha=1}^N \Delta H(\lambda(\textbf{R}_\alpha))\  ,
\end{equation}
and conserves the energy.
The compensation terms change the drift force to
\begin{equation}\label{Fdr_mix2}
\widehat {\bf  F}^{dr}_{\alpha} \equiv  -\left[  V^{AA}_\alpha     -     V^{CG}_\alpha -\left.\frac{d\ \Delta H(\lambda)}{d\lambda}\right|_{\lambda=\lambda(\textbf{R}_\alpha)}    \right]     \nabla_{\alpha}\lambda(\textbf{R}_\alpha)
\end{equation}
In the following, we relate suitable compensations to the Kirkwood's TI scheme for the free energy difference $\Delta F(\lambda)$ between a hybrid system with a position-independent coupling parameter $\lambda \le 1$ and a coarse-grained system ($\lambda = 0$) at the reference concentration $\rho^\ast$:
\begin{eqnarray}
\frac{\Delta F(\lambda)}{N} 
   &=& \frac{1}{N} \int_0^{\lambda} d\lambda'  \frac{d\ \Delta F(\lambda')}{d\lambda'}\nonumber\\
   &=& \frac{1}{N} \int_0^{\lambda} d\lambda'  \left\langle \frac{d\  H(\lambda')}{d\lambda'}\right \rangle_{\lambda'} \nonumber\\
  &=& \frac{1}{N} \int_0^{\lambda} d\lambda' \left\langle \left[ V^{AA} - V^{CG} \right]  \right\rangle_{\lambda'}
\label{KirkwoodTI}
\end{eqnarray}

\begin{figure}
\centering
\includegraphics[width=7cm]{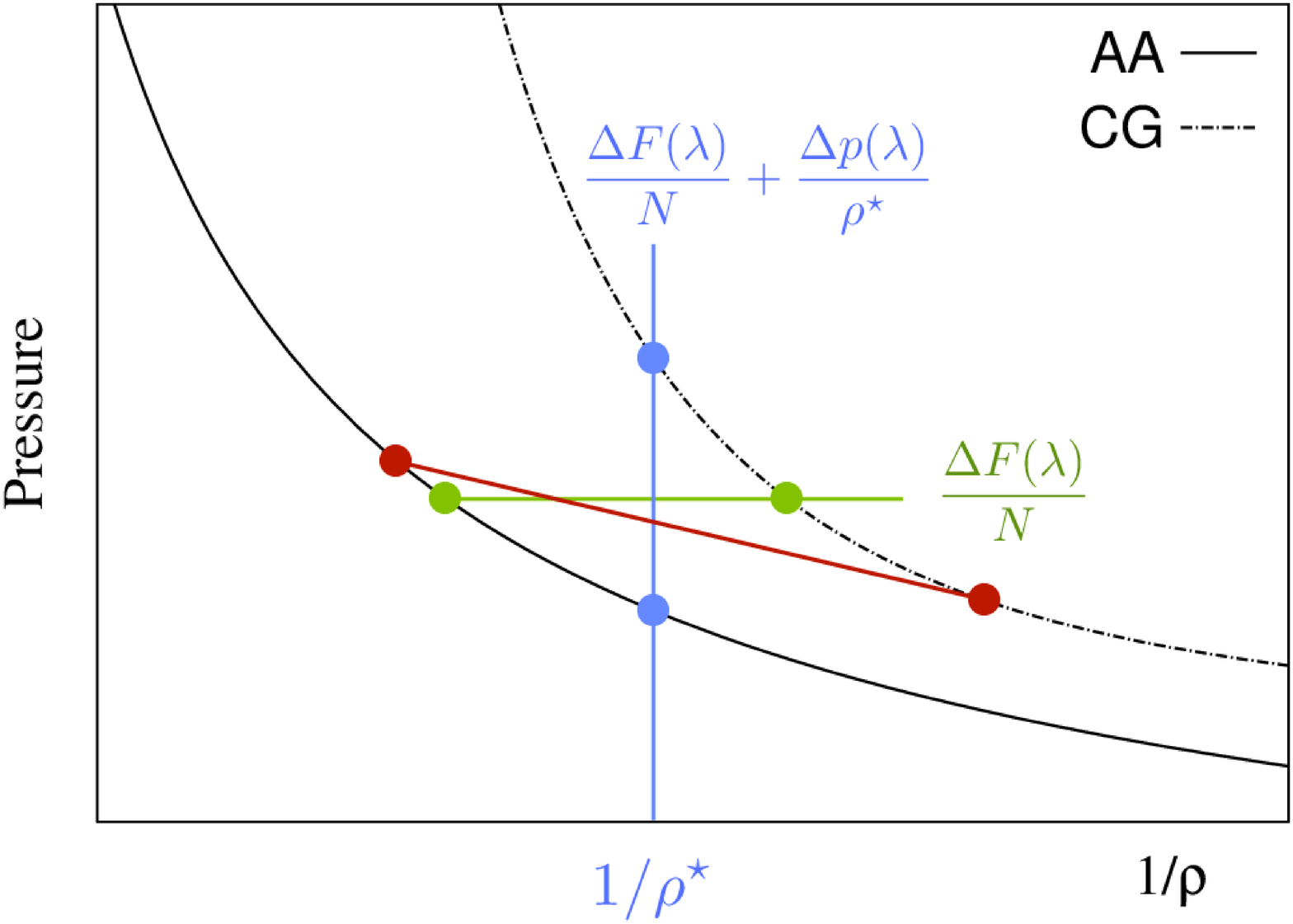}
\caption{Cartoon  illustrating  the  thermodynamics  of  H-AdResS.
 The isothermals  of AA and CG models (black solid and dashed lines) are shown.
 When no compensation term is added to the H-AdResS Hamiltonian in Eq.  (\ref{Hmix2})
 density and pressure in the two regions are different from their
 reference values (red). Applying compensation terms it is possible to maintain the coupled systems either at the same density (``constant
 density route'', blue) or at the same molecular pressure (``constant pressure route'', green).
}
\label{cartoon}
\end{figure}

Consider first a situation, where we wish to embed the AA region in a CG region with identical molecular (Virial plus kinetic) pressure.~\footnote{The molecular pressure is defined  as the sum of the kinetic and pairwise  intermolecular contributions, $P_{mol} = \rho  k_B T +\frac{1}{3V}  < \sum_{i>j}\textbf{F}_{ij} \cdot  \textbf{r}_{ij} >$;  the   single   molecule  terms   (drift   force   and  free   energy  compensations)  are  not  included.}   
To avoid the buildup of a hydrostatic pressure gradient \cite{fluid_mech} across the hybrid region, we need to assure that $\nabla p = \rho \langle \widehat {\bf F}^{dr} \rangle \equiv 0$ or
\begin{eqnarray}
\left.\frac{d\ \Delta H(\lambda)}{d\lambda}\right|_{\lambda=\lambda(\textbf{R}_\alpha)}
  \equiv \left\langle  \left[ V^{AA}_\alpha - V^{CG}_\alpha \right] \right\rangle_{{\bf R}_\alpha} 
\end{eqnarray}
If we replace the local average at each given
$\lambda=\lambda({\bf R}_{\alpha})$ by the corresponding
value in the `bulk' of a pure-$\lambda$ fluid,
$\langle \left[ V^{AA}_\alpha - V^{CG}_\alpha \right] \rangle_{{\bf 
R}_\alpha} \simeq \frac{1}{N} \left\langle \left[ V^{AA} - V^{CG} 
\right] \right\rangle_{\lambda\equiv\lambda(\textbf{R}_\alpha)}$, then 
the compensation will take the form
%
\begin{eqnarray}
\Delta H(\lambda_\alpha) = \frac{\Delta F(\lambda_\alpha)}{N}.
\label{Hmix2.1}
\end{eqnarray}
Since atomistic and coarse-grained  systems usually follow different  equations  of state ~\cite{gordon2007,wang2009comparative},
as depicted in Fig. \ref{cartoon}, the  densities of  the two regions will generally differ.
It is worth noting, though, that by adjusting the number of particles in the system one can easily
tune the particle density in the AA region to the reference value $\rho^\star$. In this case the
pressure \textit{in the entire system} would adjust to the reference value of the atomistic system.

A different compensation route has to be taken if, instead  of the  same  pressure, 
one  wants to ensure that both subsystems coexist at the  same reference density  $\rho^\star$.
In particular, the chemical potential gradient, which is
  generally established across the transition region, would have to be
  counterbalanced \cite{FritschPRL}.  This idea leads to the following
  form   of    the   compensation   term    in   Eq.   (\ref{Hmix2}),
\begin{eqnarray}\label{Hmix3}
&&\Delta H(\lambda({\bf R}_\alpha)) \equiv \Delta \mu(\lambda) = \frac{\Delta F(\lambda)}{N} + \frac{\Delta p(\lambda)}{\rho^\star},
\end{eqnarray}
where $\Delta \mu$ is the difference in chemical potential across the transition layer
and is related to the (molar) Gibbs free energy difference by
$\Delta \mu = \Delta G/N = \Delta F/N + \Delta p /\rho^{\star}$.
Again,  Kirkwood  TI  provides  a  way to  predict  $\Delta  \mu$  by
simultaneously evaluating the Helmholtz free energy difference $\Delta
F(\lambda)$  and  the   pressure  difference  $\Delta  p(\lambda)$  in
independent  simulations  of pure-$\lambda$  fluids  at the  reference
state     ($\rho^{\star}$,     $T$)     and     varying     $\lambda$.

Fig. \ref{cartoon}  graphically summarizes the  possible routes
allowed  by these  two forms  of  the free  energy compensation.   The
``pressure  route", with   $\Delta H(\lambda)  =
\Delta  F(\lambda)/N$, cancels  the extra  interface  ``pressure'' and
guarantees that mechanical equilibrium is  uniquely established
  by  inter-molecular forces  (however,  in general,  the  AA and  CG
subregions will  attain different densities).  On the other hand, in the ``density route'', the addition of
$\Delta H(\lambda)  = \Delta \mu(\lambda)$  compensates the  difference  in chemical  potential
across  the transition region  leading to  an equilibrium  state where
both subsystems  coexist at the same density,  but different molecular
pressure.
 
To test and validate the proposed compensation schemes we considered the above mentioned tetrahedral system, but we substituted the CG IBI potential with a WCA potential, which was deliberately parametrized to  give a higher
molecular pressure and $\Delta F(\lambda) <0$ 
than the atomistic  system at the same state point.
 For sake of simplicity, the same number of molecules,
total volume of the system and relative volumes of the different subregions
were used in all simulations (see also~\cite{SI}).

\begin{figure}
\centering
\includegraphics[width=7.5cm]{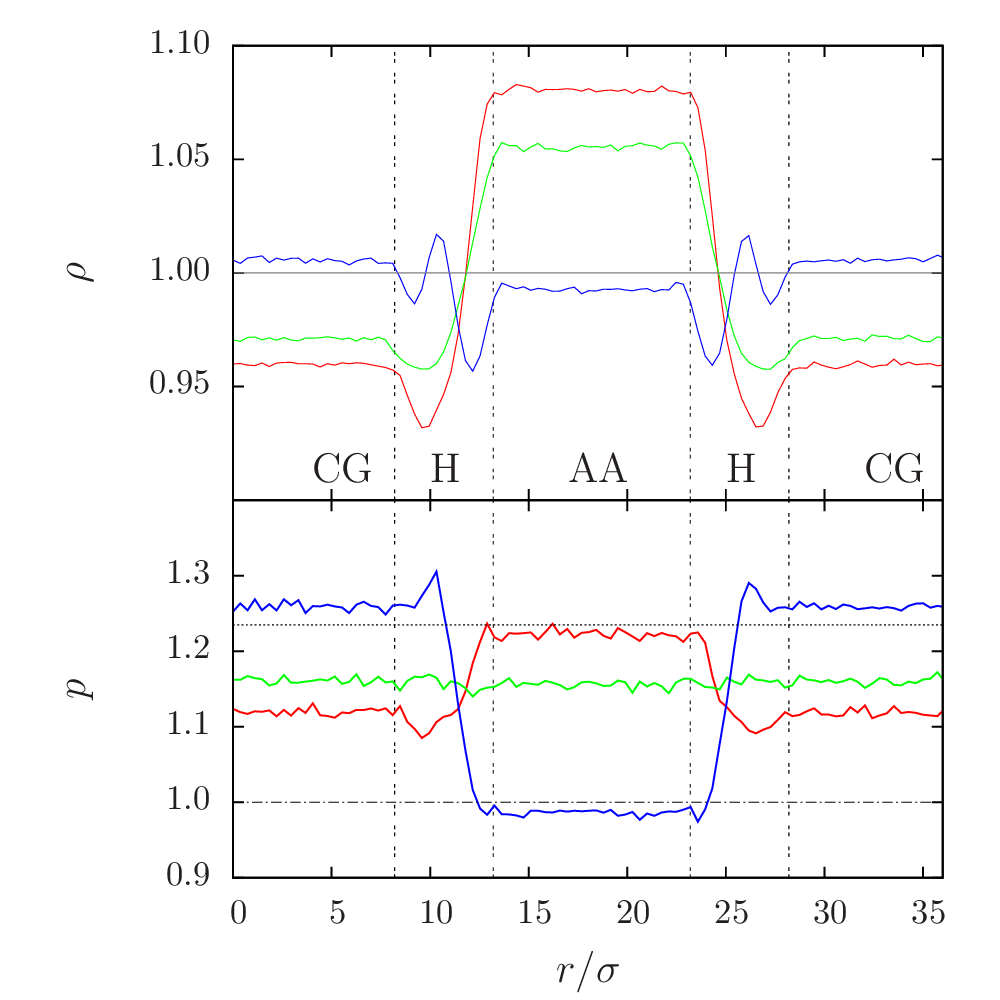}
\caption{ Plots  showing  the effect  of the  free
  energy  compensations  on the  density  profile  (upper panel)  and
  pressure  profile (lower  panel)  in a  H-AdResS  simulation with  CG
  potential having larger molecular pressure than the fully atomistic.
  The case are the same as to those presented in Fig. 2.
  Red, green and blue lines refer to simulations performed using the H-AdResS  Hamiltonian in Eq.~\ref{Hmix},  the ``constant-pressure" (Eq.~\ref{Hmix2.1}) and the "constant-density" (Eq.~\ref{Hmix3}) compensation routes, respectively.
  All  pressures  are
  normalized to the value of the all atom simulation (dash-dot
  line); the  dotted line indicates  the pressure  of the
   coarse-grained system at the reference atomistic density.}
 \label{pressdens}
\end{figure}

The red curves in Fig. \ref{pressdens} show the pressure and density profiles for
the uncorrected H-AdResS Hamiltonian, Eq. \ref{Hmix}. Both quantities exhibit
jumps in the transition regions, and in the AA region neither pressure
nor density attain the reference values. Making use of Eqs. \ref{Hmix2.1} and
\ref{Hmix3} we can compensate for the free  energy imbalance between
the AA and CG regions. The  ``constant pressure  route"
balances on average the drift force $\langle {\bf F}^{dr} \rangle$, thus
producing a flat molecular pressure profile and leading to an average
momentum conservation in the whole system; the  density  is
nonetheless  different  in  the  two  regions. In contrast,  the
``constant  density route"  levels out  the density  to  the reference
value $\rho^\star=N/V$ by  taking the pressure in the  bulky AA and CG
regions  to the  values  they have  in  the corresponding  homogeneous
simulation. The  compensation term of Eq. (\ref{Hmix3})  does not take
into  account density-density  correlations over  the  transition layer
and, as observed in Fig.  \ref{pressdens}, this produces small density
fluctuations  (of  about $\sim3\%$)  in the  transition region.  
We are currently working on a generalization of the present framework to include such correlations.
In any  case, if required, the small density fluctuations can be
removed by an iterative refinement scheme (see e.g. \cite{adresstf}).

To summarize, we have presented a method, H-AdResS, to simulate molecular liquids
with position-dependent levels of resolution. Whereas in the original AdResS scheme
the exact enforcement of Newton's 3rd law impedes a general Hamiltonian formulation~\cite{jcp,prelu},
in H-AdResS this requirement is relaxed to formulate the problem in terms of a global
Hamiltonian function. This method allows us to generate equilibrium states in any  
well defined statistical ensemble, which therefore can be sampled by either Monte Carlo
or Molecular Dynamics. In H-AdResS, the potential energies of the
molecules are weighted according to their `local nature' (atomistic, coarse-grained or
hybrid). Based on the analogy with standard Kirkwood thermodynamic integration
we have proposed two schemes to correct for the drift force appearing in the hybrid
region. Also these compensation terms are not time- or path-dependent, so that no
bookkeeping is required to enforce energy conservation; in particular, thermodynamical
equilibrium is achieved without the help of a local thermostat to remove the excess
heat produced in the hybrid region.
The pressure and density routes for free energy compensation offer a simple way
to optimize the embedding of the system as well as to modulate the thermodynamic
balance between AA and CG regions.
This new approach thus significantly widens the options to couple within a single simulation
setup representations at rather different resolution, making for a valuable tool for many problems
in soft matter science.

\begin{acknowledgments}
  The authors  are indebted with B. D\"unweg  and M. Praprotnik
    for fruitful discussions, and  acknowledge hospitality at KITP; KK
    thanks L. Delle Site and M. Praprotnik the very fruitful continued
    collaboration on the development  of the original AdResS method.
This research was supported in part by the
    National Science  Foundation under Grant No.  NSF PHY11-25915. PE
  thanks the support  of BIFI and the  Ministry of Science and
  Innovation  through project  FIS2010-22047-C05-03.  RDB also  thanks
  FIS2010-22047-C05-01 and the support  of the ``Comunidad de Madrid''
  via the project MODELICO-CM (S2009/ESP-1691).

\end{acknowledgments}


%

\end{document}